\documentclass[aps,prl,amsmath,showpacs,floatfix,superscriptaddress]{revtex4}
\usepackage{bm}
\usepackage{amssymb}
\usepackage{graphicx}
\newcommand{\dd}{\mathrm{d}}
\newcommand{\vc}[1]{\mbox{\boldmath{$#1$}}}       

\begin{document}
\title{Scaling and energy transfer in rotating turbulence}
%%\shorttitle{Scaling in rotating turbulence}
%%\author{Wolf-Christian M\"uller\inst{1} \and Mark Thiele\inst{2}}
%%\institute{
%% \inst{1} Max-Planck-Institut f\"ur Plasmaphysik - 85748 Garching, Germany \\
%% \inst{2} Universit\"at Bayreuth, Theoretische Physik II - 95440 Bayreuth, Germany
%%}
\author{Wolf-Christian M\"uller}
\affiliation{Max-Planck-Institut f\"ur Plasmaphysik, 85748 Garching, Germany}
\email{Wolf.Mueller@ipp.mpg.de}
\author{Mark Thiele}
\affiliation{Universit\"at Bayreuth, Theoretische Physik II, 95440 Bayreuth, Germany}
\email{Mark.Thiele@uni-bayreuth.de}
\affiliation{Max-Planck-Institut f\"ur Plasmaphysik, 85748 Garching, Germany}
%%\pacs{47.32.-y}{Vortex dynamics; rotating fluids}
%%\pacs{47.27.Gs}{Isotropic turbulence; homogeneous turbulence}
%%\pacs{47.27.ek}{Direct numerical simulations}
%%\maketitle 
%
\begin{abstract}
The inertial-range properties of quasi-stationary hydrodynamic
turbulence under solid-body rotation are studied via high-resolution
direct numerical simulations.  For strong rotation the nonlinear
energy cascade exhibits depletion and a pronounced anisotropy with the
energy flux proceeding mainly perpendicularly to the rotation
axis. This corresponds to a transition towards a quasi-two-dimensional
flow similar to a linear Taylor-Proudman state. In contrast to the
energy spectrum along the rotation axis which does not scale
self-similarly, the perpendicular spectrum displays an inertial range
with $k^{-2}_\perp$-behavior. A new phenomenology gives a rationale
for the observations. 
The scaling exponents $\zeta_p$ of structure functions up to order $p=8$ measured perpendicular to the rotation axis
indicate reduced intermittency with increasing
rotation rate. The proposed phenomenology is consistent with the inferred asymptotic non-intermittent behavior
$\zeta_p=p/2$.  
\end{abstract}
\maketitle

The inherent properties of turbulence in a rotating reference frame
are important for, e.g.,  the dynamics of atmosphere and oceans,
%\cite{roberts_soward:georotreview}, 
liquid planetary cores,
%\cite{busse:dynamorev}, 
%stellar convection zones,
%\cite{thompson_etal:sonnerotrev} 
and engineering problems.
%\cite{lesieur:book}.
The nonlinear spectral transfer of energy by the direct
turbulent cascade and the associated energy spectrum are particularly interesting
due to their direct connection to the 
dynamical processes governing rotating turbulence. Most of the available experimental
data
\cite{traugott:exp,wigeland:exp,ibbetson_tritton:exp,
hopfinger_browand_gagne:exp,jacquin_etal:exp} yields no conclusive
information on the expected self-similar scaling of the energy spectrum in the inertial range of scales and its
dependence on the rotation frequency $\Omega$.  Although recent
experiments \cite{baroud_etal:expspec,baroud_etal:expstruc,morize_moisy_rabaud:exp} have shed
some light on these issues, the scaling of two-point statistics in
rotating turbulence remains a controversial topic.

Direct numerical simulations
\cite{bardina_ferziger_rogallo:rotdns32,mansour_cambon_speziale:rotdns32,
hossain:rotdns32,yeung_zhou:rotdns256,smith_waleffe:rotdns200,
godeferd_lollini:rotdns48,morinishi_nakabayashi_ren:rotdns128, chen_chen_eyink_holm:rotdns128}
and
large-eddy simulations, see e.g.
\cite{squires_etal:rotles128512,bartello_metais_lesieur:rotles64,
yang_domaradzki:rotles64}, have been carried out only at low and
moderate Reynolds numbers precluding clear scaling observations.
Nevertheless, most of the cited works agree in that the
nonlinear spectral transfer of energy to smaller scales diminishes with
growing $\Omega$, accompanied by a transition of the flow towards a
quasi-two-dimensional state perpendicular to the fixed rotation axis,
$\vc{\Omega}$.  The transition manifests itself in an increasing
ratio of integral length scales parallel and perpendicular to
$\vc{\Omega}=\Omega\vc{\hat{e}}_3$, defined as $L_{i,j}=\int_0^{L_\infty}\dd
\ell \langle v_i(\vc{r})v_i(\vc{r}+\ell\vc{\hat{e}}_j)\rangle/\langle
v_i^2(\vc{r})\rangle$, $L_\infty$ representing the largest possible
distance between two points in the simulation volume and ${\ell}$
denoting the respective space increment.

This Letter presents high-resolution direct numerical simulations of
incompressible rotating homogeneous turbulence driven at largest scales and proposes a
phenomenology of the energy cascade which suggests a physical
explanation for the observed attenuation of nonlinear spectral
transfer under the influence of rotation.  In addition, the model
gives a rationale for the observed trend towards
two-dimensionality in rapidly rotating turbulence which is
corroborated by the simulations.  The scaling of 
two-point structure functions perpendicular to $\vc{\Omega}$ indicates a decreasing level
of intermittency with growing $\Omega$.% in agreement with experimental data.

Incompressible hydrodynamic turbulence under solid-body rotation is
usually described by the Navier-Stokes equations including the
Coriolis force \cite{greenspan:book} given here in dimensionless form with the
vorticity, $\vc{\omega}=\nabla\times\vc{v}$, and the non-dimensional 
kinematic viscosity $\mu$ and rotation frequency $\Omega=|\vc{\Omega}|$,
\begin{eqnarray}
\partial_t \vc{\omega} &=& \nabla\times\left(\vc{v}\times\vc{\omega}+ 
2\vc{v}\times\vc{\Omega}\right)+\mu\Delta\vc{\omega}\,, \label{ns1}\\
\nabla\cdot\vc{v}&=&0 \label{ns2}\,.
\end{eqnarray}
Equations (\ref{ns1}) and (\ref{ns2}) are numerically integrated using
an explicit trapezoidal leapfrog scheme
\cite{kurihara:trapezleapfrog} with the diffusive term included
by an integrating factor, see
e.g. \cite{meneguzzi_pouquet:convecdynamo}, while the remaining 
right-hand-side of Eq. (\ref{ns1}) is determined
pseudospectrally. The simulation volume extends $2\pi$ in each
dimension with triply periodic boundary conditions and a resolution of
$512^3$ collocation points. Aliasing errors are treated by spherical
mode truncation \cite{vincent_meneguzzi:simul}. Quasi-stationary
turbulence is generated by a forcing which freezes all modes in a sphere
of radius $k_f= 2$.

The initial state of the forced simulations is taken from non-rotating
turbulence which has been freely decaying for about one
large-eddy-turnover time, the period needed to reach the maximum of
dissipation when starting with a smooth velocity field. This initial velocity
field is characterized by an energy spectrum
$E_k\sim\exp(-k^2/k_0^2)$, $k_0=4$, and random phases.
Subsequently all Fourier modes with $k\leq 2$ are frozen. These modes sustain a gentle 
driving of the flow by nonlinear interaction with the fluctuating part of the system.
  As soon as
total energy, $E=\int_V \dd Vv^2/2$, and dissipation,
$\varepsilon=-\mu\int_V\mathrm{d}V \omega^2$, are statistically stationary with $E\simeq 1$
and $\varepsilon\simeq 0.25$ both mildly fluctuating, $\Omega$ is set
to a finite value, $5$ (system I) and $50$ (system II).

The dimensionless kinematic viscosity $\mu$ is $4\times 10^{-4}$ in
both cases.  The characteristic length $L_0$ and velocity $V_0$,
necessary for the calculation of macroscopic Rossby number,
$\mathsf{Ro}=V_0/(2\Omega L_0)$, and Reynolds number,
$\mathsf{Re}=L_0V_0/\mu$, can only be determined a posteriori in
homogeneous turbulence.  Both quantities are estimated on dimensional
grounds using $E$, $\varepsilon$, and $\Omega$ as $L_0\sim E/(\Omega
\varepsilon)^{1/2}$ and $V_0\sim E^{1/2}$.  Hence the Rossby and Reynolds
number given in this paper are defined as
$\mathsf{Ro}=\sqrt{\varepsilon/(4\Omega E)}$ and
$\mathsf{Re}=\sqrt{E^3/(\Omega\varepsilon)}/\mu$, respectively.  Note
that another common estimate of the Rossby number is
$\mathsf{Ro}^*=4\mathsf{Ro}^2$.

After the sudden onset of rotation, $E$ displays a sharp drop of about $20\%$
(case I) and $13\%$ (case II) with a subsequent remount that  
levels off near the previous state. 
The dissipation rate $\varepsilon$ follows
the general dynamics of energy, but does not increase again with
$\varepsilon\simeq 0.05$ in both simulations. The observations can be
understood by the rotation-induced depletion of the spectral energy transfer 
which causes a transient until forcing and cascade have reached a new
equilibrium. The observed behavior does not differ qualitatively if the rotation 
is ramped up (as has been checked by test computations). 

\begin{figure}
\centerline{\includegraphics[width=0.5\textwidth]{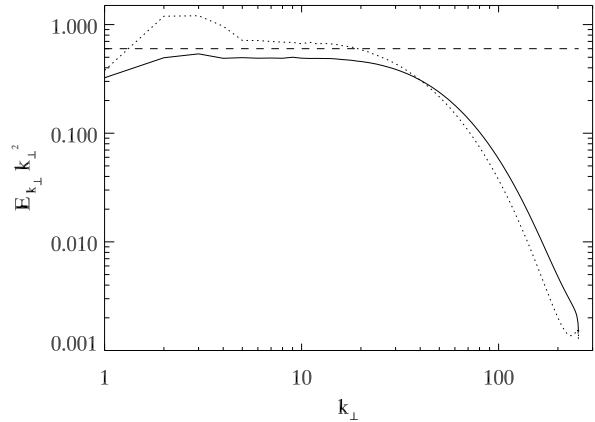}}
\caption{One-dimensional energy spectra perpendicular to the rotation
axis for different rotation frequencies. The spectra are time-averaged
over 6 large-eddy turnover times and compensated with $k_\perp^2$ (solid line: $\Omega=5$, dotted line: $\Omega=50$).}
\label{f1} \end{figure}

The following $15$ (I) and $10$ (II) large-eddy turnover times of
statistically stationary rotating turbulence are characterized by
$\mathsf{Ro}\simeq5.3\times 10^{-2}$, $\mathsf{Re}\simeq 4000$ (I) and
$\mathsf{Ro}\simeq1.3\times 10^{-2}$, $\mathsf{Re}\simeq 2340$ (II).
Perpendicular one-dimensional energy spectra, $E_{k_\perp}=\int\dd
k_\parallel\int\dd k' |v_{\vc{\scriptscriptstyle k}}|^2/2$ with $k_\perp\perp\vc{\Omega}$,
$k_\parallel\parallel \vc{\Omega}$, and $k'$ perpendicular to $k_\perp$ and $k_\parallel$, 
are shown in Fig. \ref{f1}. The spectrum of system I displays a scaling
range for $4\lesssim k_\perp \lesssim 20$ while in simulation II the 
dissipation region is starting at smaller $k$ (see below). In addition, the
spectrum of simulation II exhibits a bump around $k_f$ where the forcing region descends 
into the freely evolving range of scales. This effect which is caused by the simple forcing
scheme does not seem to significantly perturb the flow beyond $k\approx
5$. It is therefore tolerable at the chosen numerical resolution.

Although the inertial range in simulation II is shorter than in
simulation I the perpendicular one-dimensional energy spectra in both
cases exhibit scaling, $E_{k_\perp}\sim k_\perp^{-2}$, in agreement
with %experimental data \cite{baroud_etal:expspec}, 
direct numerical
simulations at moderate Reynolds-number \cite{yeung_zhou:rotdns256}
and \cite{smith_waleffe:rotdns200}(for $k>k_f$), and shell-model
calculations \cite{hattori_rubinstein_ishizawa:rotshell}. A formal
analysis of the energy flux in helical mode decomposition
\cite{canuto_dubovikov:formal} leads to the same result as well as
dimensional analysis of the energy flux terms which occur in
quasi-normal closure theories \cite{zhou:dimanal,mahalov_zhou:dimanal}
when assuming $\tau_*\sim \tau_\Omega \sim \Omega^{-1}$ for the
relaxation timescale of nonlinear interactions, $\tau_*$.
The
observed spectra are also in accord with one of several isotropic
scalings proposed in weak-turbulence theory
\cite{galtier:rotweak,cambon_rubinstein_godeferd:rotweak}.
However, this approach is
only valid in the asymptotic limit $\tau_* \gg\tau_\Omega$ implying
$k\ll k_\Omega$ (see below) which requires very strong rotation or an
extremely broad inertial range and therefore is beyond the scope of the
simulations considered here.  In \cite{smith_waleffe:rotdns200}
$k^{-3}$-scaling is found at small wavenumbers in turbulence with
medium-scale forcing. The observation is explained dimensionally by
the missing explicit $\varepsilon$-dependence of the spectrum at large
scales (also cf. \cite{smith_lee:reducedmodel} using a method based on  
reduced sets of nonlinear interactions).
Recent experimental results \cite{morize_moisy_rabaud:exp}
suggest an energy scaling exponent $\approx -2.5$ for case II with the
micro Rossby number of
$\mathsf{Ro}_\omega=\langle\omega_3^2\rangle^{1/2}/(2\Omega)\simeq
0.08$ and an exponent of $\approx -1.7$ for case I with
$\mathsf{Ro}_\omega\simeq 0.7$. This disagreement is probably due to
the different way of turbulence generation. While in the simulations
there is a continuous large-scale forcing, turbulence in the experiment 
is exited initially and then subject to decay under rotation. In addition the 
experimental Rossby numbers are significantly larger than in our computations. 
The configuration is,  therefore, not directly comparable to
the simulations described here.
This is also true for the experiment reported in \cite{baroud_etal:expspec}
although the same scaling $\sim k^{-2}$ is observed there (however in an inverse energy cascade).
It should be noted that the $k_\parallel$-dependence of the perpendicular energy spectrum (not shown)
in our simulations confirms 
the expected concentration of energy around the plane $k_\parallel=0$ 
\cite{cambon_lacquin:edqnm3,cambon_mansour_godeferd:etransfer}. 
Furthermore, the perpendicular energy spectra
at specific fixed $k_\parallel$ do not show clear spectral scaling. This is only seen in the sum $E_{k_\perp}$.

\begin{figure}
\centerline{\includegraphics[width=0.5\textwidth]{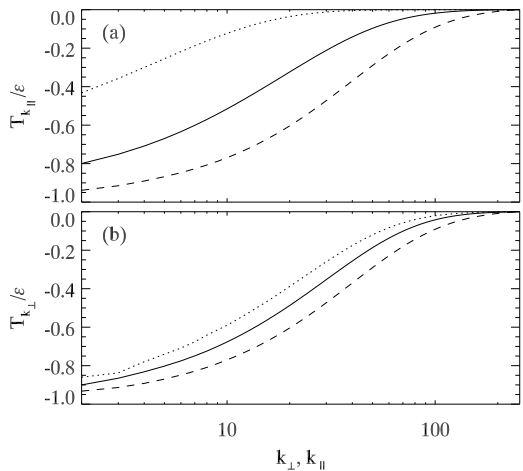}}
\caption{Normalized and time-averaged nonlinear energy fluxes (a) parallel and
(b) perpendicular to the rotation axis with $\Omega=0$ (dashed line),
$\Omega=5$ (solid line), and $\Omega=50$ (dotted line).} \label{f2} \end{figure}

The energy spectra taken parallel to $\vc{\Omega}$ do not exhibit
distinct scaling ranges. This is in accord with the strong
rotation-induced decrease of the axis-parallel nonlinear energy flux
(cf. Fig. \ref{f2}), $T_{k_\parallel}=\int_0^{k_\parallel}\dd
k_\parallel \int\dd k'\int\dd k_\perp \left(i\vc{\omega}^*\cdot
(\vc{k}\times\widetilde{\left[\vc{v}\times\vc{\omega}\right]}_{\vc{\scriptscriptstyle
k}}) +c.c.\right)/\vc{k}^2$, with $\widetilde{[\bullet]}$ denoting
Fourier transformation and $T_{k_\perp}$ defined analogously.  Note
that the Coriolis force has no direct effect on the kinetic energy
since it is oriented perpendicularly to $\vc{v}$. It does, however, modify
the nonlinear interactions leading to depletion and anisotropy of the
energy cascade, cf. also
\cite{cambon_lacquin:edqnm3,cambon_mansour_godeferd:etransfer}, which
is apparent when regarding Fig. \ref{f2}. All normalized transfer
functions for $\Omega=0$, $5$, and $50$ are negative throughout
indicating a direct energy cascade towards small-scales. They show a
damping of the energy flux at all scales with increasing $\Omega$.
The cascade depletion is much stronger in $T_{k_\parallel}$ than in
$T_{k_\perp}$. Consequently, the cascade becomes highly anisotropic in
the case of strong rotation which naturally leads to a dynamical
two-dimensionalization of the flow.
The transition towards 2D also manifests itself in the increasing ratio of parallel
to perpendicular integral length scales with growing $\Omega$ where strong rotation,
$\Omega=50$, yields $L_{i,3}$ up to a factor $2.5$ larger than
$L_{i,1}$ and $L_{i,2}$ for $i=1,2,3$. This trend is also seen
in visualizations of the velocity field (not shown).  The
perpendicular longitudinal and lateral correlation lengths exhibit
moderate systematic differences due to the forcing scheme which is not
fully isotropic. This, however, does not cloud the main trend of
two-dimensionalization in planes perpendicular to $\vc{\Omega}$.

The obtained results can be understood in the framework of a
phenomenology of the energy cascade in rotating turbulence which is
set out in the following. While inertial waves present in rotating turbulence 
are undoubtedly an important dynamical process, the proposed picture of the energy cascade 
is based on the presumed energetic dominance of convective motions.

The wavenumber range strongly influenced
by rotation lies above the scale at which the advection nonlinearity
in Eq. (\ref{ns1}) is roughly equal to the Coriolis force,
$v_\ell^2/\ell \sim \Omega v_\ell$ where $v_\ell$ denotes an
isotropically estimated velocity fluctuation at scale $\ell\sim
k^{-1}$.  Since non-rotating hydrodynamic turbulence exhibits
Kolmogorov inertial-range scaling \cite{kolmogorov:k41a},
$v_\ell\sim(\varepsilon\ell)^{1/3}$, one obtains the well-known
rotation wavenumber \cite{zeman:rotscale,canuto_dubovikov:formal},
$k_\Omega=(\Omega^3/\varepsilon)^{1/2}$, below which the energy
cascade is modified by the Coriolis force acting in planes perpendicular
to $\vc{\Omega}$.

The nonlinear energy transfer in those planes can be estimated
dimensionally as $\varepsilon\sim v^2_\xi/\tau_\mathrm{tr}$ with
the characteristic velocity fluctuation $v_\xi$ in the axis-perpendicular
plane at scale $\xi\sim k_\perp^{-1}$.  Due to the lack of
an inertial range in the direction parallel to $\vc{\Omega}$, a consequence of
the quasi-2D-state, the
approximation $E(\ell)\sim v^2_\xi$ suffices for the following
scaling predictions.  In isotropic non-rotating hydrodynamic
turbulence fluid particles which belong to a turbulent structure at
scale $\xi\sim\ell$ follow trajectories of length $\xi$ for a
turnover time, $\tau_\mathrm{tr}\sim
\tau_\mathrm{NL}\sim\xi/v_\xi$, in the course of the energy
cascade.  When neglecting all nonlinearities fluid particles in
rotating flows are displaced from the no-rotation trajectories in
axis-perpendicular directions due to the Coriolis force and follow
circular orbits of radius $r\sim v_\xi\tau_\Omega$ which close
after $\tau_\Omega\sim \Omega^{-1}$. The nonlinear terms in
Eq. (\ref{ns1}) which dominate turbulence dynamics cause a strong
deformation of the circular orbits.  The deformed circles which we
regard as abstract entities whose shapes can be left unspecified will
be referred to as \lq arcs\rq . Typically these do not close in
themselves after $\tau_\Omega$ but lead to an effective
axis-perpendicular displacement $\sim r$ of the fluid particles. It is
easy to verify that in the rotation dominated range of scales, $k <
k_\Omega$, the displacement $r$ is always smaller then $\xi$,
requiring the fluid particles to perform $\xi/r$ \lq arc steps\rq\
to cover the distance $\xi$ and to complete the cascade
trajectory. In fact though, the direction in which a fluid particle is
deflected by an arc movement is quasi-stochastic. Consequently,
analogous to a random walk process $(\xi/r)^2$ arc steps are necessary
to cover the distance $\xi$.  Therefore, the nonlinear transfer
time in the axis perpendicular direction is given by
\begin{equation}\tau_\mathrm{tr}\sim\left(\frac{\tau_\mathrm{NL}}{
\tau_\Omega}\right)^2\tau_\Omega\,.\label{rottrans}\end{equation} This
result has also been obtained by formal analysis of the nonlinear
energy flux \cite{canuto_dubovikov:formal} and in weak-turbulence
theory \cite{galtier:rotweak}.  We note in passing that 
the rotation-modified transfer time
$\tau_\mathrm{tr}$ is larger than $\tau_\mathrm{NL}$ with
$\tau_{NL}\sim\tau_\mathrm{tr}$ at $k_\Omega$. 
From the previous it is clear that for $k\ll k_\Omega$ the turbulent energy cascade is highly anisotropic and
progresses
mainly in the direction perpendicular to the rotation axis in accord with 
the dynamic Taylor-Proudman theorem, see e.g. \cite{mahalov_zhou:dimanal,chen_chen_eyink_holm:rotdns128}.
The definition of $\tau_\Omega$ used in this paper differs from the one
known from weak turbulence theory which involves the ratio
$k_\perp/k_\parallel$ (see, e.g., \cite{galtier:rotweak}). However, both definitions approach each other when 
$k_\perp/k_\parallel$ for the turbulent fluctuations does not depart strongly from unity as is the case here.

With relation (\ref{rottrans}) the dimensional estimate of the
nonlinear energy flux is obtained as $\varepsilon\sim
v_\xi^4/(\Omega\xi^2)$. Assuming 
$\varepsilon=\mathrm{const.}$ throughout the inertial range yields a
scaling law for the velocity fluctuations perpendicular to $\vc{\Omega}$,
\begin{equation} v_\xi\sim
(\Omega\varepsilon)^{1/4}\xi^{1/2}\,,\label{rotscal}
\end{equation} corresponding to the observed scaling of the energy spectrum $\sim
k_\perp^{-2}$ \cite{zhou:dimanal,canuto_dubovikov:formal}.

The rotation-dominated range of scales is limited from below by the
wavenumber $\min(k_\Omega,k^\Omega_d)$ where $k^\Omega_d$ indicates approximate equality of 
nonlinear and dissipative energy-fluxes, $\varepsilon\sim
\mu\xi^{-2}v_\xi^2$. Together with relation (\ref{rotscal})
this yields $k_d^\Omega\sim \mu^{-1}(\varepsilon/\Omega)^{1/2}\sim
k_\Omega\frac{\varepsilon}{\mu\Omega^2}$. A different approach
\cite{canuto_dubovikov:formal} leads to the same result.  
Since $k_\Omega$ grows and $k^\Omega_d$ diminishes with increasing $\Omega$, 
there exists a critical
rotation frequency for which the rotation-dominated regime has its largest extent, 
i.e. $R=\varepsilon/(\mu\Omega^2)=1$. Higher rotation rates 
lead to a reduction of this range since $R<1$.
Here,
for case I: $k_\Omega\simeq 50$, $k_d^\Omega\simeq 250$, $k_d\simeq
167$, $R\simeq 5$ and for case II: $k_\Omega\simeq 1581$, $k_d^\Omega\simeq 79$,
$k_d\simeq 167$, $R\simeq 0.05$ which explains the shorter inertial range of
$E_{k_\perp}$ in case II. The Kolmogorov dissipation wavenumber $k_d\sim(\varepsilon/\mu^3)^{1/4}$
roughly identifies the scale at which dissipation begins to dominate over the nonlinear energy flux.

\begin{figure}
\centerline{\includegraphics[width=0.5\textwidth]{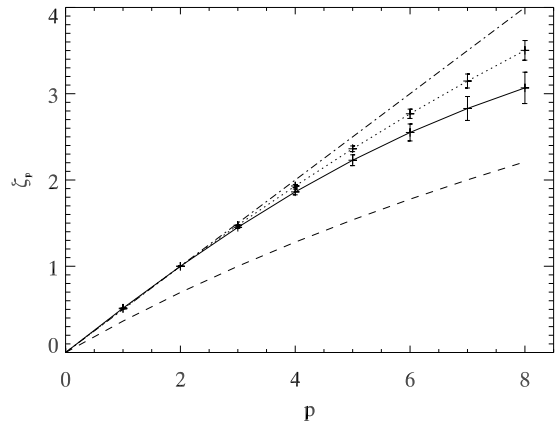}}
\caption{Axis-perpendicular structure function scaling exponents $\zeta_p$ of velocity up
to order 8 for moderate and strong rotation. Solid line: $\Omega=5$, dotted line: $\Omega=50$, dashed line:
She-L\'ev\^eque intermittency model, dash-dotted line: non-intermittent
scaling, $\zeta_p=p/2$.}  \label{f3} \end{figure}

At the given spatial resolution scaling exponents of the
axis-perpendicular longitudinal velocity structure functions,
$\langle|[\vc{v}(\vc{r})-\vc{v}(\vc{r}+\vc{\xi})]\cdot
\vc{\xi}/\xi|^p \rangle\sim\xi^{\zeta_p}$ 
can be determined for both systems up to order 8. However, the small inertial
range for $\Omega=50$ necessitates use of the extended self-similarity property (ESS) \cite{benzi:ess}.
The relative exponents $\zeta_p/\zeta_2$ obtained via ESS
approximately coincide with the $\zeta_p$ since
relation (\ref{rotscal}) (in the non-intermittent limit) and the numerical result $E_{k_\perp}\sim k_\perp^{-2}=k_\perp^{-(\zeta_2+1)}$ 
suggest $\zeta_2 \approx 1$. 
The results shown in Fig. \ref{f3} are a sign of a 
gradual transition from the intermittent non-rotating case
(represented by the She-L\'ev\^eque formula \cite{she_leveque:model},
$\zeta_p= p/9+2[1-(2/3)^{p/3}]$) towards a strongly rotating flow with weak intermittency.
The observed reduction of intermittency is in accord with the expected transition
from strong fluid turbulence to weak inertial wave turbulence 
(see, e.g., \cite{biven_connaughton_newell:weakbreakdown}).  
The observations coincide with
experimental findings given in \cite{baroud_etal:expstruc} although in this experiment the flow is driven at
small scales and in contrast to our simulations exhibits an inverse cascade of energy.

In summary, high-resolution direct numerical simulations of
incompressible hydrodynamic turbulence driven at largest scales under moderate and strong
rotation corroborate a proposed phenomenology of rotating turbulence
which gives a simple rationale for the overall weakening of the energy cascade and the
trend towards two-dimensional dynamics in rotating turbulent flows.
Higher-order structure function scalings show a transition towards a non-intermittent state
perpendicular to the rotation axis as known from 2D turbulence.% and in agreement with experimental data.
\begin{acknowledgments} 
It is a pleasure to thank Friedrich Busse
and Dieter Biskamp for helpful discussions.
\end{acknowledgments} 
\bibliography{/home/Wolf/Texte/Bibliographien/Turbulence,/home/Wolf/Texte/Bibliographien/RotatingTurb,/home/Wolf/Texte/Bibliographien/Dynamo,/home/Wolf/Texte/Bibliographien/Convection}

\end{document}